# Photonic glass-ceramics: consolidated outcomes and prospects

Brigitte Boulard[1], Tran T. T. Van[2], Anna Łukowiak[3], Adel Bouajaj[4], Rogéria Rocha Gonçalves[5], Andrea Chiappini[6], Alessandro Chiasera[6], Wilfried Blanc[7], Alicia Duran[8], Sylvia Turrell[9], Francesco Prudenzano[10], Francesco Scotognella[11], Roberta Ramponi[11], Marian Marciniak[12], Giancarlo C. Righini[13,14], Maurizio Ferrari[6,13,*]

[1] Institut des Molécules et Matériaux du Mans, UMR 6283, Equipe Fluorures, Université du Maine, Av. Olivier Messiaen, 72085 Le Mans cedex 09, France.
[2] University of Science Ho Chi Minh City, 227 Nguyen Van Cu, Dist.5, HCM Vietnam
[3] Institute of Low Temperature and Structure Research, PAS, ul. Okolna 2, 50-950 Wroclaw, Poland
[4] Laboratory of innovative technologies, LTI, ENSA–Tangier, University Abdelmalek Essaâdi, Tangier, Morocco.
[5] Departamento de Química, Faculdade de Filosofia, Ciências e Letras de Ribeirão Preto, Universidade de São Paulo - Av. Bandeirantes, 3900, CEP 14040-901, Ribeirão Preto/SP, Brazil
[6] CNR-IFN, CSMFO Lab., Via alla Cascata 56/c, Povo, 38123 Trento, Italy.
[7] Université Nice Sophia Antipolis, CNRS LPMC, UMR 7336, 06100 Nice, France.
[8] Instituto de Ceramica y Vidrio (CSIC), C/Kelsen 5, Campus de Cantoblanco, 28049 Madrid, Spain.
[9] LASIR (CNRS, UMR 8516) and CERLA, Université Lille 1, 59650 Villeneuve d'Ascq, France.
[10] Politecnico di Bari, DEI, Via E. Orabona 4, Bari, 70125, Italy.
[11] IFN-CNR and Department of Physics, Politecnico di Milano, p.zza Leonardo da Vinci 32, 20133 Milano, Italy
[12] National Institute of Telecommunications, 1 Szachowa Street, 04 894 Warsaw, Poland.
[13] Centro di Studi e Ricerche "Enrico Fermi", Piazza del Viminale 2, 00184 Roma, Italy.
[14] MipLAB. IFAC - CNR, Via Madonna del Piano 10, 50019 Sesto Fiorentino, Italy.

## ABSTRACT

Transparent glass-ceramics are nanocomposite materials which offer specific characteristics of capital importance in photonics. This kind of two-phase materials is constituted by nanocrystals embedded in a glass matrix and the respective composition and volume fractions of crystalline and amorphous phase determine the properties of the glass-ceramic. Among these properties transparency is crucial, in particular when confined structures, such as dielectric optical waveguides and optical fibers, are considered, and the number of papers devoted to this topic is continuously increasing. Another important point is the role of the nanocrystals when activated by luminescent species, as rare earth ions, and their effect on the spectroscopic properties of the glass-ceramic. The presence of the crystalline environment around the rare earth ion allows high absorption and emission cross sections, reduction of the non-radiative relaxation thanks to the lower phonon cut-off energy, and tailoring of the ion-ion interaction by the control of the rare earth ion partition. This last point is crucial and still object of intense experimental and theoretical studies. The composition of the glass matrix also impacts the properties of the rare earth ions located in nanoparticles. Moreover, some kinds of nanocrystals can play as effective rare earth sensitizers. Fabrication, assessment and application of glass-ceramic photonic systems, especially waveguides, deserve an appropriate discussion which is the aim of this paper, focused on luminescent glass-ceramics. In this work, a brief historical review, consolidated results and recent advances in this important scientific and technological area will be presented, and some perspectives will be outlined.

*maurizio.ferrari@ifn.cnr.it Tel: +39 0461 314250 Fax: +39 0461 314875

# INTRODUCTION

Glass photonics is pervasive in a huge number of human activities and drive the research in the field of enabling technologies. The fruitful exploitation of glass photonics is not restricted only to the area of Information and Communication Technology. Many other photonic devices, with a large spectrum of applications covering Health and Biology, Structural Engineering, and Environment Monitoring Systems, have been developed during the last years. Glass materials and photonic structures are the cornerstones of scientific and technological building in integrated optics. Photonic glasses, optical glass waveguides, planar light integrated circuits, waveguide gratings and arrays, functionalized waveguides, photonic crystal heterostructures, and hybrid microresonators are some examples of glass-based integrated optical devices that play a significant role in optical communication, sensing, biophotonics, processing, and computing. This cross-disciplinary approach leads to constructed luminescent structures that can perform sensing and functionalized structures to successfully address socioeconomic challenges, such as security, cost-effective healthcare, energy savings, efficient and clean industrial production, environmental protection, and fast and efficient communications. Photonics, with its pervasiveness, has already been identified as an enabling technology, and through advanced research in glass-based integrated optics systems, photonics can contribute to finding new technical solutions to still unsolved problems, and pave the way to applications not yet imagined.

Among the different glass-based systems, transparent glass-ceramics are nanocomposite materials which offer specific characteristics of capital importance in photonics. These two-phase materials are constituted by nanocrystals embedded in a glass matrix and the respective composition and volume fractions of crystalline and amorphous phase determine the properties of the glass-ceramic. Transparency is crucial property, in particular for dielectric optical waveguides and optical fibers. Another important point is the effect of the nanocrystals activated by rare earth ions on the spectroscopic properties. Fabrication, assessment and application of glass-ceramic photonic systems, deserve an appropriate discussion which is the aim of this paper, focused on luminescent glass-ceramics.

# A SHORT HISTORY OF GLASS-CERAMICS

The scientific and technological activity involving the development of materials at nano-micro scale and related converging technologies allows progress in the conception, design, and realisation of systems and devices with substantially improved performance and significant scientific results. This activity is coming from far both in terms of time and motivation. Table I presents some milestones concerning a brief history of glass-ceramic materials. René-Antoine Ferchault de Réaumur, the brilliant physicist and naturalist, was also in charge of the manufacture of porcelain. He tried to produce a porcelain which could compete with the porcelain of China, without being able to fully solve the problem but creating that opaque glass that bears the name of "porcelain Reaumur" (1739) [1]. We have to wait until the year 1952 to appreciate the invention of glass-ceramics thanks to the experiment performed by the Corning Glass Works scientist Stanley Donald Stookey. In the description of the experiment we read that the furnace containing the glass accidentally overheated. Instead a melted glass Stookey observed a white material with the shape of the original glass. He recognized the obtained material as a specific new material that we know as "glass-ceramic" and was initially described as a polycrystalline ceramic materials, formed through the controlled nucleation and crystallization of glass, where the amount of residual glassy phase is usually less than 50% [2]. The commercial outcome of this material was the immediate exploitation of its unique and amazing thermal and mechanical properties leading to the cookware pieces still now known as Pyroflam and Vision products. These strong and thermal shock resistant glass-ceramics have monopolized the research for a long time until the interest in transparent glass-ceramics has made its way thanks to both fundamental research and the required novel application in optics. Is only in 1968 that Schott AG produced a transparent glass-ceramic called Zerodur employed for a number of very large telescope mirrors and also used as sample reference to test the validity of innovative glass structural models [3]. The first application of transparent glass ceramics in the field of photonics appears in 1995 with the paper of Tick et al. concerning the exploitation of transparent glass ceramics for optical amplifiers operating at 1300 nm [4]. In the abstract of the paper authors say:" The properties of an oxyfluoride glass ceramic that possesses high transparency after ceramming are described. Approximately 25 vol% of this material is comprised of cubic, fluoride nanocrystals and the remainder is a predominantly oxide glass. When doped with $Pr^{3+}$, the fluorescence lifetime at 1300 nm is longer than in a fluorozirconate glass, suggesting that a significant fraction of the rare-earth dopant is preferentially partitioned into the fluoride crystal phase. This material has the added advantage of being compatible with ambient air processing". It's worthy to note that this paper contains the more significant topics that will be developed in the next 20 years in the field of photonic glass ceramics, i.e. (i) the fundamental role of oxyfluoride as parent glass; (ii) the role of the nanocrystals in reducing the concentration quenching; (iii) the role of nanocrystals with low cut-off energy in reducing non-radiative transitions [4]. After this work we can find in literature a huge amount of papers regarding rare earth activated oxyfluoride based transparent glass ceramics [5-8].

Table 1. A brief history of glass-ceramics

| Milestone | Date |
|---|---|
| René-Antoine Ferchault de Réaumur produced polycrystalline materials from silica-soda–lime | *1739* |
| Stanley Donald Stookey - Corning's scientists - discovered glass-ceramics | *1952* |
| VISIONS and PYROFLAM Registered Corning Glass Works Trademark | *1958* |
| ZERODUR by Schott | *1968* |
| Transparent glass ceramics for 1300 nm amplifier applications | *1995* |
| The relationship between structure and transparency in glass-ceramic materials | *2000* |

However, the main result in photonic glass ceramics has been obtained in waveguiding application. A rare earth-activated glass-ceramic waveguide constitutes a potential significant system to behave as an effective optical medium for light propagation and luminescence enhancement [9]. Looking at this specific application the problem of the transparency becomes crucial and in fact n 1998 Tick published a paper with the explicit title "Are low-loss glass-ceramic optical waveguides possible?" [10]. The answer was positive and he concluded the paper suggesting that the minimum transmission loss limit of the investigated effective medium glass ceramics is in the order of tens of decibels per kilometre (dB/km), once all of the impurities are eliminated. In the same paper some general criteria for light propagation are given. These criteria concern nanocrystals size, narrow particle-size distribution, inter-particle spacing, and clustering. In 2000 Tick, Borrelli and Reaney published another significant paper discussing the relationship between structure and transparency in glass-ceramic materials into a near single-mode optical waveguide fiber geometry [11]. In this significant paper authors investigate the transparency analysing the composition dependence of the lattice parameters and refractive index of the nanocrystals, their size distribution, the particle separation and the actual crystalline volume fraction [11]. A detailed scattering analysis was performed using two different approaches. One is as a contiguous two-phase system with the scattering characterized by a correlation function. The other is a particle approach, where one uses an effective medium scattering model. The authors conclude that although the latter predicts the correct order of magnitude of the scattering, it cannot distinguish the scattering behaviour as a function of ceramming conditions used. This point is still of crucial interest and several efforts have been done to achieve general rules to correlate scattering attenuation with appropriate ceramming protocols. From the point of view of simulation many techniques have been developed for analyzing scattering problems and several interesting solutions have been discussed [12]. Moving to fabrication, although a general behaviour is not yet determined excellent results have been obtained for specific compositions [13-18] and we will present some examples in this paper.

## GEO$_2$ TRANSPARENT GLASS CERAMIC PLANAR WAVEGUIDES

Recently effort is devoted on growth, characterization on GeO$_2$ based system for his employment for applications on various modern fields such as electronics and photonics. GeO$_2$ exhibits, in fact, many interesting properties that make this material suitable for applications in optical, electronic and optoelectronic devices [19, 20, 22]. It is of strategic importance to develop fabrication protocols allowing obtain GeO$_2$ transparent glass ceramic planar waveguides exhibiting low attenuation coefficients and simultaneously embed GeO$_2$ nanocrystals with a specific phase [22]. Among the various techniques can be used to fabricate these particular kinds of nanostructured systems such as sol-gel techniques with top-down and bottom-up approaches, and physical vapor deposition, we have demonstrate that RF-sputtering (RFS) is a suitable route to fabricate optical planar waveguides and photonic microcavities operating in the visible and NIR regions [20, 23]. In order to produce transparent glass-ceramic, a common procedure employ heat treatments of the systems, but lately efforts was performed for the development of alternative treatments such as laser annealing (LA) that presents advantages in terms of temporal and spatial annealing control in respect to the conventional thermal annealing [20, 24]. The CO$_2$ laser annealing has been successfully used to reduce scattering losses in different kinds of amorphous optical planar waveguides [20], and for the fabrication of glass-ceramic coating and glass-ceramic waveguides with low attenuation coefficient [20]. Anyway it is of crucial importance a precise characterization of the structure of the samples to assess the effect of the LA process and tailor the irradiation protocols. Moreover, it is well known that important structural modification, leading to artificial results, can be induced in GeO2 based materials by electron irradiation [20]. For these reasons, information on the structural properties of the waveguides before and after the laser annealing process are obtained by various techniques and in particular from positron annihilation spectroscopy (PAS), specifically by the Doppler broadening spectroscopy (DBS). PAS is a well-known non-destructive spectroscopy technique employed to investigate materials structure [25]. Experimental details could be found in [19].

In Table 2 are reported the optical parameters for the samples as prepared and after $CO_2$ laser irradiation for 2h. The waveguides have a thickness of about 1μm and support one mode at 1319 nm and 1542 nm. Before and after LA the refractive indices measured in TE and TM polarization modes are equal within the experimental uncertainty, so the birefringence can be considered negligible. Comparing the refractive indices in the $GeO_2$ waveguides before and after LA we observe an increase of about 0.04 with the irradiation at all the wavelengths.

Table 2. Optical parameters for the samples as prepared and after $CO_2$ laser irradiation for 2h.

| Laser wavelength (nm) | | Refractive Index | | thickness (μm) | Attenuation coefficient (dB/cm) |
|---|---|---|---|---|---|
| | | TE | TM | | |
| 632 | before irradiation | 1.614±0.001 | 1.616±0.001 | 1.1±0.1 | 1.9±0.2 |
| | after irradiation | 1.652±0.001 | 1.653±0.001 | 1.1±0.1 | 1.1±0.2 |
| 1319 | before irradiation | 1.590±0.01 | 1.590±0.01 | 1.0±0.1 | 1.4±0.2 |
| | after irradiation | 1.631±0.01 | 1.634±0.01 | 1.0±0.1 | 0.7±0.2 |
| 1542 | before irradiation | 1.585±0.01 | 1.585±0.01 | 1.0±0.1 | 0.9±0.2 |
| | after irradiation | 1.623±0.01 | 1.624±0.01 | 1.0±0.1 | 0.5±0.2 |

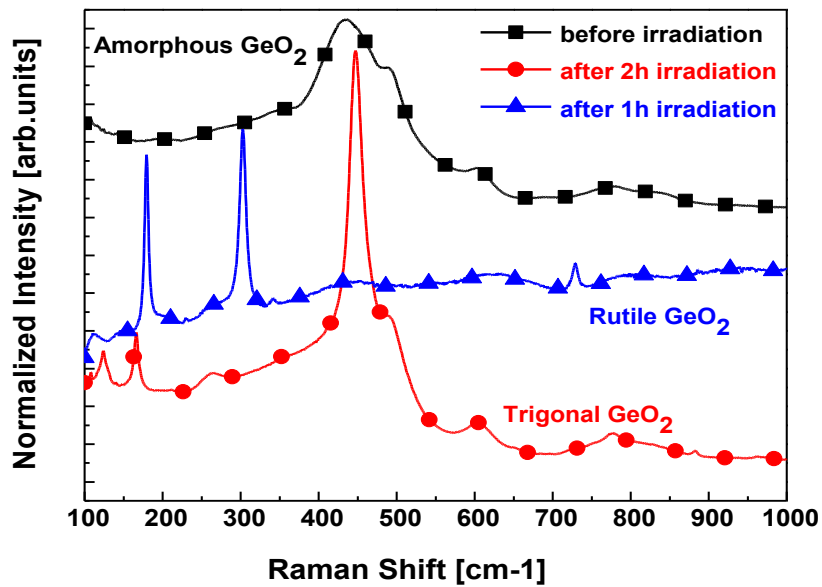

Figure 1 Micro Raman measurements carried out at room temperature for $GeO_2$ planar waveguide.

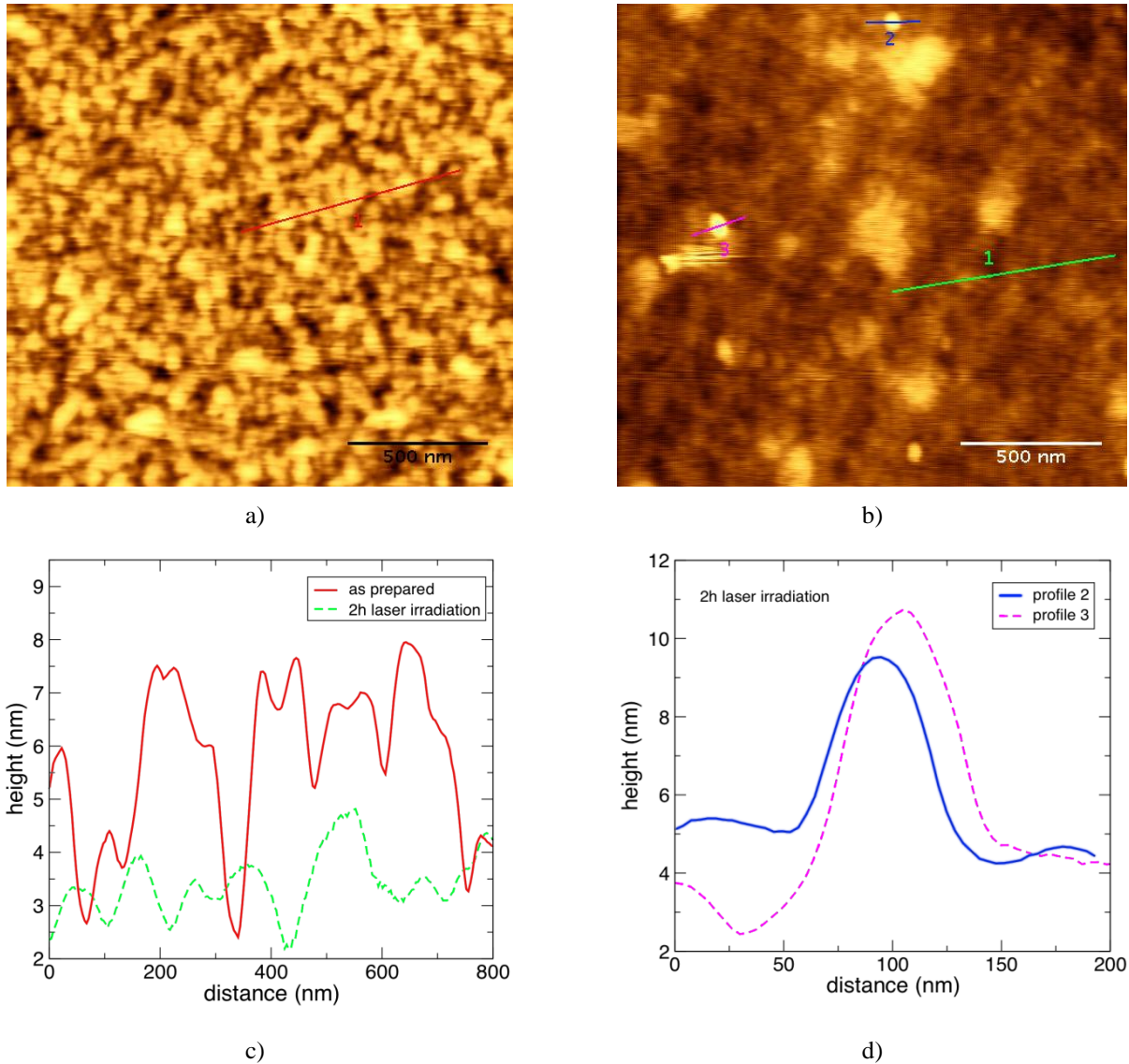

Figure 2. AFM images of representative 1.8 x 1.8 μm$^2$ areas of the samples in the conditions: (a) as prepared and (b) after 2h of $CO_2$ laser irradiation. Z scale 10 nm. In (c) a comparison of typical height profiles in these two conditions, corresponding to sections 1 in panels (a) and (b), showing a decrease in roughness after irradiation is presented. (d) profiles of the nanometric structures found after irradiation - sections 2 and 3 in panel (b).

The laser annealing allowed significant reduction of the attenuation coefficients. In fact, we observe an attenuation coefficient at 1542 nm of 0.5 dB/cm for the irradiated system while for the system before LA is 0.9 dB/cm. As shown in Table 2, the reduction of the attenuation coefficient with the CO2 laser irradiation is present at all the considered wavelengths. The decrease in the total attenuation coefficient for the CO2 laser irradiated systems, that take in account of all the contribution to the losses such as volume and surface scattering and absorption, has been attributed by Dutta et al. [24] to the reduction of the surface roughness

The optical parameters measured after 1h of LA are equivalent to those reported for 2h. EDXS analysis was used to monitor the chemical composition of the sample, as deposited and after LA. EDXS confirm that the composition of the sample is not affected by the laser irradiation and the correct stoichiometry between germanium and oxygen is always present.

In figure 1, the Micro Raman spectra measured at room temperature for GeO2 planar waveguides before and after LA at different times are reported. The Raman spectrum of the as prepared GeO2 waveguide before irradiation is typical of a GeO2 amorphous system. The amorphous nature of the GeO2 film is clearly pointed out by the broad peak at 440 cm-1 [20]. After 1 h of CO2 laser irradiation, the Raman spectra indicate that in the amorphous GeO2 matrix there is the presence of a rutile-like GeO2 crystalline phase [20]. After 2 h of CO2 laser irradiation, the Raman spectrum shows that in the amorphous system there is also the presence of a trigonal GeO2 crystal phase [20].

AFM images of representative 1.8 x 1.8 µm2 areas for the as prepared sample and for that CO2 laser irradiated for 2h are reported in figure 2 (a) and 2 (b) respectively. AFM analysis put in evidence a roughness of 1.05 ± 0.05 nm in the as prepared sample and a roughness of 0.80 ± 0.02 nm for the waveguide after 2h of CO2 laser irradiation. This result correlates with the reduction of the attenuation coefficient with the laser annealing and confirms surface morphology as an important source of losses. Moreover, on the surface of the sample, the AFM image of figure 2 (b) shows the presence of nanometric structures with sizes ranging from 10 to 50 nm (see also figure 2 (c)). These structures are assigned to GeO2 nanocrystals. It is worth noting that, although the presence of these scattering points are expected to increase the attenuation coefficient by increasing the scattering losses of the waveguide, the protocol here developed for the CO2 irradiation allows reducing the attenuation coefficient by decreasing the surface scattering losses [24].

Table 2. $S_n$ values characterizing each layer in the as prepared and the two irradiated samples, as obtained by fitting the positron depth profiles. The density value of the $GeO_2$ film was used as a guess parameter into the frame of the fitting procedure. The boundary depths are measured from the surface of the sample.

| Sample | layer I $\rho = 3.15$ g/cm$^3$ | | layer II $\rho = 3.15$ g/cm$^3$ | | layer III $\rho = 3.15$ g/cm$^3$ | | bulk $\rho = 2.1$ g/cm$^3$ |
|---|---|---|---|---|---|---|---|
| | $S_n$ | depth (nm) | $S_n$ | depth (nm) | $S_n$ | depth (nm) | $S_n$ |
| before irradiation | 1.006 | 11 ± 2 | 0.952 | 983 ± 20 | -- | -- | 1.000 |
| after 1h irradiation | 1.003 | 2 ± 0.2 | 0.946 | 320 ± 25 | 0.966 | 983 | 1.000 |
| after 2h irradiation | 0.980 | 5 ± 0.1 | 0.950 | 175 ± 32 | 0.960 | 983 | 1.000 |

In the as prepared film, positrons detect a first thin superficial layer of about 10 nm which is characterized by a very high $S_n$ value of 1.006. This thin layer is followed by another $GeO_2$ layer (labelled layer II) uniform up to the silica substrate and characterized by a $S_n = 0.952$. With the laser irradiation, the superficial layer reduces to a half of its initial thickness, and the uniform second layer observed in the as prepared film splits now into two regions, named layers II and III, respectively. The first region is structurally similar to the second layer of the non-irradiated sample having about the same $S_n$ value (~ 0.950). In the sample irradiated during 1 h the thickness of the second layer is around 300 nm; and in the case of the sample irradiated 2 h, this thickness decreases up to approximately 170 nm. On the other hand, the second region (i.e., layer III) is characterized by a higher $S_n$ value (~ 0.960) than the previous one and it extends up to the silica interface. The detection by DBS of a thin superficial layer that decreases its thickness with the laser irradiation treatment, well-correlates with the results obtained by AFM. Probably, this layer is highly defected due to their $S_n$ values by which it is characterized (higher than the $S_n$ values found in the bulk of the films). Both, the decrease in roughness and thickness of the defected superficial layer could be linked to the decrement of the attenuation coefficient. From the results reported in Table 2, it can be seen that the $S_n$ values obtained for layer III are slightly higher than those of layer II, indicating a structural change of the $GeO_2$ samples. Besides, the thickness of this modified layer is larger in the sample annealed for a longer irradiation time. In the case of layer II, for the laser treated films $S_n$ values are almost equal to that not treated sample. Moreover we can note that: a) the bulk of the film interested by the structural change increases by increasing the irradiation time from 1 to 2 h; b) the change starts from the film/substrate interface moving towards the surface of the films. Raman spectroscopy reveals the presence of different $GeO_2$ phases. The rutile $GeO_2$ phase present after 1 h of irradiation exhibits a higher refractive index than that of the trigonal phase observed after 2 h of irradiation [20]. However, DBS results would allow inferring that the lack of substantial changes in the macroscopic refractive index measured after 1 and 2 h of irradiation, respectively, could be attributed to a balance between the contributions of the different phases to the modified matrixes. The behavior described in point b) could be assigned to the heating of the substrate by the $CO_2$ laser inducing a progressive modification of the film from the substrate to the surface. The $SiO_2$ matrix, in fact, due to the lattice vibration in the region of 940 $cm^{-1}$, presents a higher absorbance in the region of the $CO_2$ laser line with respect to that of the $GeO_2$ system [20]. For that reason the $CO_2$ radiation is mainly absorbed by the $SiO_2$ substrate that induces the change in the $GeO_2$ film.

## $Er^{3+}$-ACTIVATED $SnO_2$ SOL-GEL-DERIVED GLASS-CERAMICS: AN ENERGY TRANSFER CASE STUDY

During the last years several work has been performed in order to develop energy transfer based systems for photoluminescence application. In this scientific area we have up- and down-converters for planar waveguides for integrated optics [17], visible laser light sources and solar cells efficiency enhancement [18,26], specific nanostructured system used as sensitizers of rare earth luminescence [27]. Recently, a detailed work regarding the effect of the local environment on the spectroscopic properties of rare-earth-activated tin dioxide glass ceramic was presented [28, 29].

Tin dioxide, which is a high-refractive index semiconductor ($E_g$ = 3.6 eV at 300K), and is stable both chemically and mechanically, is an excellent choice for the crystalline component of the system. As it is also transparent through the visible and infrared regions, it covers the emission range for active ions like erbium. Compatibility is enhanced by the fact that tin oxide nano-crystals can be excited by a broad range of UV wavelengths, thus with the use of broad band arc lamps, while $Er^{3+}$ ions have very narrow excitation peaks.

Studies on silica-tin bulk glass ceramics have shown that the concentration of $SnO_2$ has a definite effect on the structure

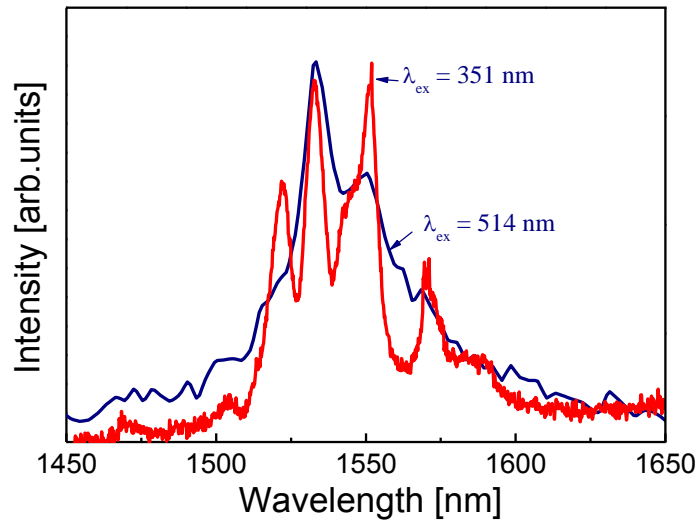

Figure 3: Photoluminescence spectra of a $(100-30)SiO_2$ - $30SnO_2$ waveguide, doped with 0.5 mol% $Er^{3+}$ and heat-treated at 1200°C, upon excitations at 351 and 514.5 nm

of the $SiO_2$ glass phase [30,31]. In addition, while rare earth ion concentrations as high as 1 mol% can be doped into the glass ceramic, $Er^{3+}$ ions are found mainly in the glass phase, which results in high levels of quenching effects. In order to investigate any possible energy transfer between erbium ions and $SnO_2$ nanocrystals, the environment of the $Er^{3+}$ ions was studied via infrared photoluminescence measurements which were recorded using 351 nm and 514.5 nm excitation wavelengths. These two lines correspond to the excitation of the $SnO_2$ crystals and to the excitation of the $^4I_{15/2} \rightarrow {}^2H_{11/2}$ transition of $Er^{3+}$ ions across their bandgap, respectively.

Figure 3 shows the photoluminescence spectra of the $^4I_{13/2} \rightarrow {}^4I_{15/2}$ transition of $Er^{3+}$ ion for the samples containing 30 mol% $SnO_2$, doped with 0.5 mol% $Er^{3+}$ and heat-treated at 1200°C. Under excitation at 514.5 nm, one obtains an emission spectrum characteristic of $Er^{3+}$ ions in an amorphous medium, with a broad band centered around 1535 nm. However, excitation at 351 nm results in a completely different spectrum, in which the presence of narrow bands at 1521, 1531, 1549 and 1571 nm are assigned to the Stark transitions between $^4I_{13/2}$ and $^4I_{15/2}$ levels . When the erbium ion occupies the $Sn^{4+}$ sites in the cassiterite structure, the electrostatic field of the surrounding environment of ions removes degeneracy of the free ion J levels and makes them split into doubly degenerated stark level [28,29]. This emission results from energy transfer between $SnO_2$ nanoparticles and the rare-earth ions by Forster's mechanism. Accordingly, this spectrum indicates that most part of $Er^{3+}$ ions are embedded in $SnO_2$ nanocrystals. Consideration of these spectral features indicates that the $Er^{3+}$ ions exist in two types of sites: those embedded in tin oxide nanocrystals and those within the amorphous silica matrix. The influence of the incorporation of erbium in the tin dioxide nanoparticles on luminescence were studied by the photoluminescence measurements of films with different $Er^{3+}$ concentrations. With increasing erbium concentration, the near-infrared emission intensity at 1540 nm is steadily enhanced. The PL intensity of the 2 mol% Er-doped sample is more than three orders of magnitude greater than that of the 0.5 mol% Er-doped one. This behavior indicates that quenching of luminescence did not occur in the 2 mol% erbium doped systems.

## CONCLUSIONS

Transparent glass ceramic photonic systems have demonstrated to be crucial for several applications and especially in integrated optics. The importance of these nanocomposite materials is evident when luminescence efficiency and attenuation losses are the more important requirements. Concerning the role of nanocrystals as rare earth sensitizers, $SnO_2$ nanocrystals constitutes a textbook example. $SnO_2$ is a wide band-gap (Eg = 3.6 eV at 300 K) n-type semiconductor and exhibits a maximum phonon energy of about 630 $cm^{-1}$. The rare earth ion can be incorporated in the $SnO_2$ nanocrystal and substitutes for the $Sn^{4+}$ ions in the cassiterite-structured $SnO_2$ nanocrystals. Among the different fabrication techniques we have mentioned the laser annealing used in the case of $GeO_2$ transparent glass ceramic planar waveguides. The $GeO_2$ planar waveguide after 2h of $CO_2$ laser irradiation exhibits an increase of 0.04 in the refractive index, measured at 1542 nm. We demonstrate that the laser annealing significantly reduce propagation loss in $GeO_2$ planar waveguides that laser annealing decreases the surface roughness. From another side Raman results have shown that laser treatment for 1 h and 2 h results in realization of structures at nanometric scale on the surface of the samples with dimension from 10 to 50 nm attributed to the presence of $GeO_2$ nanocrystals. These behaviors evidence the role of the laser heat treatment on the glass inducing a progressive modification of the films.

We can therefore conclude that the prospects of glass photonic are still very good, with continuously growing applications in the area of integrated optics, lasing, lightning, frequency converters and sensing. Optimization of the synthesis processes of glasses tailored for the specific application and of the confined structures fabrication technologies may guarantee that new record performances of rare earth doped glass based photonic devices will be achieved. In respect to the development of photonic glass ceramics the immediate problems to solve are: (i) reproducible fabrication protocols; (ii) modelling of transparency constraints including disorder; (iii) enhance solubility for the rare earth ions.

## ACKNOWLEDGEMENTS

This research was performed in the framework of the projects CNR-PAS (2014-2016), CNR-CNRST (2014-2016), COST ACTION MP1401, Premiale Centro Fermi PIHH, MaDEleNA PAT (2013-2016). R.R. Gonçalves and M. Ferrari acknowledge Brazilian Scientific Mobility Program "Ciências sem Fronteiras".

Let us to thank a number of young colleagues who are closely collaborating to this topic. Some of them are working at their PhD thesis: Simone Normani, Anna Piotrowska, Iustyna Vasilchenko, other are highly appreciated for their invaluable technical support: Cristina Armellini, Alessandro Carpentiero, Maurizio Mazzola, Stefano Varas.

## REFERENCES

[1] Baumé A., "Chimie expérimentale et raisonnée", vol. 3, P. Franç. Didot le jeune, Ed.,Paris,1773.
[2] Holand W., Beall G.H., Glass Ceramic Technology, The Am. Ceramic Society, Westerville, OH, 2002.
[3] Champagnon B., Boukenter A., Duval E.:, "Early stages of nucleation of Zerodur glass: Very low frequency Raman scattering and small angle X-ray scattering investigations", J. Non-Cryst. Solids, 94 pp. 216-221 (1987).

[4] Tick P.A., Borrelli N.F., Cornelius L.K., Newhouse M.A., "Transparent glass ceramics for 1300 nm amplifier applications", J. Appl. Phys., 78 pp. 6367-6374 (1995).
[5] Tikhomirov V.K., Furniss D., Seddon A.B., Reaney I.M., Beggiora M., Ferrari M., Montagna M., Rolli R.: "Fabrication and characterization of nano-scale, $Er^{3+}$-doped, ultra-transparent oxy-fluoride glass-ceramics", Appl. Phys. Lett., 81 pp. 1937-1939 (2002).
[6] Beggiora M., Reaney I.M., Islam M.S., "Structure of the nanocrystals in oxyfluoride glass ceramics", Appl. Phys. Lett., 83 pp. 467-469 (2003).
[7] Mattarelli M., Montagna M., Moser E., Chiasera A., Tikhomirov V., Seddon A.B., Chaussedent S., Nunzi Conti G., Pelli S., Righini G.C., Zampedri L., Ferrari M., "$Tm^{3+}$-activated transparent oxy-fluoride glass ceramics: structural and spectroscopic properties", J. Non-Cryst. Solids, 345&346 pp. 354-358 (2004).
[8] de Pablos-Martín A., Ristic D., Bhattacharyya S., Höche T., Mather G.C., Ramírez M.O., Soria S., Ferrari M., Righini G.C., Bausá L.E., Durán A., Pascual M.J., "Effects of $Tm^{3+}$ addition on the crystallization of $LaF_3$ nanocrystals in oxyfluoride glasses: optical characterization and up-conversion", J. Am. Ceram. Soc., 96 pp.447-457 (2013).
[9] Guddala S., Alombert-Goget G., Armellini C., Chiappini A., Chiasera A., Ferrari M., Mazzola M., Berneschi S., Righini G.C., Moser E., Boulard B., Duverger Arfuso C., Bhaktha S.N.B., Turrell S., Narayana Rao D.,. Speranza G. "Glass-ceramic waveguides: Fabrication and properties", in Proc. ICTON 2010, Munich, Germany, July 2010, paper We.C2.1.
[10] Tick P.A.:, "Are low-loss glass-ceramic optical waveguides possible?, Opt. Lett., 23 pp. 1904-1905 (1998).
[11] Tick P.A., Borrelli N.F., Reaney I.M., "The relationship between structure and transparency in glass-ceramic materials", Opt. Mat., 15 pp. 81-91 (2000).
[12] Doicu A., Wriedt T., Eremin Y.A., "Light Scattering by Systems of Particles", Springer-Verlag, 2006.
[13] Ferrari M. and Righini G.C., Rare-earth-doped glasses for integrated optical amplifiers", in Physics and Chemistry of Rare-Earth Ions Doped Glasses Materials Science Foundations (monograph series), Vol. 46-47 (2008), Trans Tech Publishers (ttp), Switzerland, Editors: N. Sooraj Hussain and J.D. Santos.
[14] Péron O., Boulard B., Jestin Y., Ferrari M., Duverger-Arfuso C., Kodjikian S., and Gao Y., "Erbium doped fluoride glass-ceramics waveguides fabricated by PVD", J. Non-Cryst. Solids, 354 pp. 3586-3591 (2008).
[15] Alombert-Goget G., Armellini C., Bhaktha S.N.B., Boulard B., Chiappini A.,. Chiasera A., Duverger-Arfuso C., Féron P., Ferrari M., Gonçalves R.R., Jestin Y., Minati L., Monteil A., Moser E., Nunzi Conti G., Osellame R., Pelli S., Quandt A., Ramponi R., Rao D. N., Righini G.C., Speranza G., Vishnubhatla K.C., "Silica-hafnia-based photonic systems", The Mediterranean Journal of Electronics and Communications, 6 pp.8-17 (2010).
[16] Shivakiran Bhaktha B.N., Berneschi S., Nunzi Conti G., Righini G.C.,Chiappini A., Chiasera A., Ferrari M., Turrell S., "Spatially localized UV-induced crystallization of $SnO_2$ in photorefractive $SiO_2$-$SnO_2$ thin film", Proc. of SPIE, 7719. pp. 77191B-1/5 (2010)..
[17] Chiasera A., Alombert-Goget G., Ferrari M., Berneschi S., Pelli S., Boulard B., Duverger Arfuso C.R"are earth activated glass-ceramic in planar format", Opt. Eng., 50 pp. 071105-1/10 (2011).
[18] Dieudonné B., Boulard B., Alombert-Goget G., Chiasera A., Gao Y., Kodjikian S., Ferrari M., "Up- and down-conversion in $Yb^{3+}$-$Pr^{3+}$ co-doped fluoride glasses and glass ceramics", J. Non-Cryst. Solids 377 pp. 105-109 (2013).
[19] Chiasera A., Macchi C., Mariazzi S., Valligatla S., Varas S., Mazzola M., Bazzanella N., Lunelli L., Pederzolli C., Rao D. N., Righini G. C., Somoza A., Brusa R., Ferrari M., "$GeO_2$ glass ceramic planar waveguides fabricated by RF-sputtering", Proc. of SPIE 8982 pp. 89820D-1/12 (2014).
[20] Chiasera A., Macchi C., Mariazzi S., Valligatla S., Lunelli L., Pederzolli C., Rao D.N., Somoza A., Brusa R.S., Ferrari M., "$CO_2$ laser irradiation of $GeO_2$ planar waveguide fabricated by rf-sputtering" Optical Materials Express 3 pp.1561-1570 (2013).
[21] Nunzi Conti G., Berneschi S., Brenci M., Pelli S., Sebastiani S., Righini G. C., Tosello C., Chiasera A., and Ferrari M.,"UV photoimprinting of channel waveguides on active $SiO_2$–$GeO_2$ sputtered thin films," Appl. Phys. Lett. 89 pp. 121102-1/3 (2006).
[22] Prakapenk V. P., Shen G., Dubrovinsky L. S., Rivers M. L., and Sutton S. R., "High pressure induced phase transformation of $SiO_2$ and $GeO_2$: difference and similarity" J. Phys. Chem. Solids 65 pp.1537-1545 (2004).
[23] Valligatla S., Chiasera A., Varas S., Bazzanella N., Rao D.N., Righini G.C., and Ferrari M., "High quality factor 1-D $Er^{3+}$-activated dielectric microcavity fabricated by rf-sputtering," Opt. Express 20 pp. 21214–21222 (2012).

[24] Dutta S., Jackson H.E., Boyd J.T., Davis R.L., and Hickernell F.S., "$CO_2$ laser annealing of $Si_3N_4$, $Nb_2O_5$ and $Ta_2O_5$ thin-film optical waveguides to achieve scattering loss reduction," IEEE J. Quantum Electron. 18 pp. 800–806 (1982).

[25] Macchi C., Mariazzi S., Karwasz G. P., Brusa R. S., Folegati P., Frabboni S., and Ottaviani G., "Single-crystal silicon coimplanted by helium and hydrogen: evolution of decorated vacancylike defects with thermal treatments," Phys. Rev. B 74 pp.174120-1/12 (2006).

[26] Alombert-Goget G., Ristic D., Dieudonné B., Moser E., Varas S., Berneschi S., Ivanda M., Monteil A., Arfuso Duverger C., Righini G. C., Boulard B., Ferrari M." Rare-earth-activated glasses for solar energy conversion", Proc. ICTON 2011, Stockholm, Sweden, June 2011, paper We.B6.1.

[27] Alombert-Goget G., Armellini C., Berneschi S., Bhaktha S.N.B., Boulard B., Chiappini A., Chiasera A., Duverger-Arfuso C., Féron P., Ferrari M., Jestin Y., Minati L., Monteil A., Moser E., Nunzi Conti G., Pelli S., Prudenzano F., Righini G.C., Speranza G., "Glass-Based Erbium Activated Micro-Nano Photonic Structures", Proc. ICTON 2009, Azores, Portugal, June 2009, paper We.A5.4.

[28] Tran Van T.T., Turrell S., Capoen B., Le Van Hieu, Ferrari M., Ristic D., Boussekey L., Kinowski C., "Environment segregation of $Er^{3+}$-emission in bulk sol-gel-derived $SiO_2$-$SnO_2$ glass ceramics", Journal of Materials Science 49 pp. 8226-8233 (2014).

[29] Tran Van T. T., Turrell S., Capoen, B. Lam Vinh Q., Cristini-Robbe O., Bouazaoui M., d'Acapito F., Ferrari M., Ristic D., Lukowiak A., Almeida R., Santos L. and Kinowski C., "Erbium-doped tin-silicate sol-gel-derived glass-ceramic thin films: Effect of environment segregation on the $Er^{3+}$ emission", Science of Advanced Materials 7 pp. 301-308 (2015).

[30] Van Tran T. T., Si Bui T., Turrell S., Capoen B., Roussel P., Bouazaoui M., Ferrari M., Cristini O. and Kinowski C., "Controlled $SnO_2$ nanocrystal growth in $SiO_2$-$SnO_2$ glass-ceramic monoliths",Journal of Raman Spectroscopy 43 pp. 869-875 (2012).

[31] Van Tran T., Turrell S., Eddafi M., Capoen B., Bouazaoui M., Roussel P., Berneschi S., Righini G.C., Ferrari M., Bhaktha S. N. B., Cristini O. and Kinowski C., "Investigations of the effects of the growth of $SnO_2$ nanoparticles on the structural properties of glass-ceramic planar waveguides using Raman and FTIR spectroscopies", Journal of Molecular Structure 976 pp 314-319 (2010).